\documentclass[twocolumn,preprintnumbers,amsmath,amssymb,pra,showpacs]{revtex4}

\usepackage{theorem}
\usepackage{enumerate}
\usepackage{amsmath}
\usepackage{amssymb}
\usepackage{graphicx}
\usepackage{epstopdf}
\usepackage{color}
\usepackage{euscript}
\usepackage{float}

\theorembodyfont{\rmfamily}
\newtheorem{Theorem}{\textit{Theorem} }
\newtheorem{Proposition}{Proposition}
\newtheorem{Definition}{\textit{Definition} }
\newtheorem{Corollary}{\textit{Corollary}}

\newtheorem{Proof}{\textit{Proof}}

\newcommand{\lv}{\left \vert}
\newcommand{\rv}{\right \vert}
\newcommand{\la}{\left \langle}
\newcommand{\ra}{\right \rangle}
\newcommand{\ket}[1]{\lv #1 \ra}
\newcommand{\bra}[1]{\la #1 \rv}
\newcommand{\braket}[2]{\langle #1 \vert #2 \rangle}
\newcommand{\ketbra}[2]{\lv #1 \rangle \langle #2 \rv}

\newcommand{\expectation}[1]{\mathbf{E}[#1]}

\newcommand{\tr}{\mathrm{Tr}}
\newcommand{\EMW}{E_{\mathrm{MW}}}

\begin{document}

\title{Phase-random states: ensembles of states with fixed amplitudes and uniformly distributed phases in a fixed basis}
\author{Yoshifumi Nakata,$^1$ Peter S. Turner,$^1$ and Mio Murao$^{1,2}$}
\affiliation{$^1$Department of Physics, Graduate School of Science, University of Tokyo, Tokyo 113-0033, Japan\\
$^2$Institute for Nano Quantum Information Electronics, University of Tokyo, Tokyo 153-8505, Japan}

 \begin{abstract}
Motivated by studies of typical properties of quantum states in statistical mechanics, we introduce phase-random states, an ensemble of pure states with fixed amplitudes and uniformly distributed phases in a fixed basis.
We first give a sufficient condition for canonical states to typically appear in subsystems of phase-random states,
which reveals a trade-off relation between the initial state in the bounded energy subspace and the energy eigenstates that define that subspace.
We then investigate the simulatability of phase-random states, which is directly related to that of time evolution in closed systems, by studying their entanglement properties. 
We find that starting from a separable state, time evolutions under Hamiltonians composed of only separable eigenstates generate extremely high entanglement and are difficult to simulate with matrix product states.  
We also show that random quantum circuits consisting of only two-qubit diagonal unitaries can generate an ensemble with the same average entanglement as phase-random states.
\end{abstract}

\date{\today}


\maketitle

\section{Introduction}
One of the goals of quantum many-body physics is to be able to compute properties such as expectation values of observables, entanglement
and explicit state descriptions for physical systems composed of many particles. These properties depend upon the parameters of the quantum states involved, and calculations are made difficult by the fact that the number of these parameters grows exponentially with the number of particles. One way around this problem is to consider {\it ensembles} of states as opposed to individual states, as many of those parameters are then averaged out.

The most natural and well-studied ensemble of states is that of  {\it random states},  the set of pure states in Hilbert space selected randomly from the unitarily invariant distribution, used in many areas of quantum physics and quantum information science \cite{LPSW2009, HP2007,BHL2005}.
The entanglement of random states is an area of particular recent interest, where it has been shown that in large systems, the {\it average} amount of entanglement is nearly maximal according to several measures \cite{L1978}.  
In the context of quantum statistical mechanics, the high entanglement of random states has been shown to lead to reduced states that are thermal, under their restriction to a subspace constrained by the amount of total energy \cite{GLTZ2006}.  

Symmetry is often used as a constraint, however there are other ways to place useful restrictions on an ensemble.
If states are restricted to a certain subspace of Hilbert space, an ensemble of states could be described by random states in that subspace. 
In this paper, we address another type of restriction, which leads to ensembles of states in a {\it subset} of a Hilbert space.
A simple example of a subset of states that is not a subspace is the set of product states, as they are not closed under superpositions. 
A more interesting example is a state evolving under a time-independent Hamiltonian. 
Since the time evolution changes only the phases of the expansion coefficients in the Hamiltonian's eigenbasis, states reached during the time evolution form a subset rather than a subspace.
Our aim is to study random states in such a subset, namely, an ensemble of states where the randomness is restricted to the phase of the complex expansion coefficients in a given basis, which we call \emph{phase-random states}.

Phase-random states are closely connected to studies of typical properties in statistical mechanics \cite{GLTZ2010, LPSW2009}. 
States in which the phases are the pertinent degree of freedom appear in quantum information theory such as in instantaneously quantum polynomial time (IQP) circuits \cite{SB2009}, and locally maximally entanglable (LME) states \cite{KK2009}. 
Phase-random states also describe situations where the accessible information is limited to the amplitudes in a certain basis,
for instance, those where a unique rank-$1$ measurement is allowed.

Motivated by these considerations, we investigate statistical properties of phase-random states, which clearly depend on the amplitudes of the coefficients as well as the expansion basis.  
We first demonstrate the thermalization of their reduced states, which implies potential uses of phase-random ensembles to realize thermal states in subsystems. 
Then, regarding phase-random states as typical states during a Hamiltonian dynamics, we discuss {\it simulating} them with matrix product states (MPSs) \cite{ECP2010} by deriving the average amount of entanglement of phase-random states. 
Moreover, applications of phase-random states in IQP circuits and as LME states lead us to develop a scheme for generating an ensemble of states simulating the entanglement of phase-random states by a quantum circuit composed of only diagonal two qubit unitaries, which we call a {\it phase-random circuit}.

The paper is organized as follows.
In Sec.~\ref{Sec:PRS}, we define phase-random states and show how they can be used to study thermalization in statistical mechanics.
In Sec.~\ref{Sec:Ent}, we derive explicit formula for the average amount of entanglement of phase-random states.
Using the formula, we investigate the simulatability of Hamiltonian dynamics by MPSs in Sec.~\ref{Sec:Sim}.
Finally, we introduce and analyze phase-random circuits in Sec.~\ref{Sec:PRC}.

\section{Phase-random states and thermalization}  \label{Sec:PRS}
Given a Hilbert space $\mathcal{H}$, we denote an ensemble of pure states $\ket{\psi} \in \mathcal{H}$ distributed according to some measure d$\mu$ by $\Upsilon = \{ \ket{\psi} \}_{{\rm d}\mu}$.  
The ensemble of random states is written $\Upsilon_{\rm rand} = \{ \ket{\psi} \}_{{\rm d}\psi}$, where $\ket{\psi}$ is an arbitrary state and d$\psi$ is the unitarily invariant normalized Haar measure. 	 
For a Hilbert space of $N$ qubits, consider states of the form
\begin{equation}
\ket{\phi} =  \sum_{n=1}^{2^N} r_n e^{i \varphi_n} \ket{u_n}, \notag
\end{equation}
with both the amplitudes $\{r_n | \sum_n r_n^2 =1, \; 0 \leq r_n \leq 1 \}$ and orthonormal basis $\{ \ket{u_n} \}$ fixed.  
By phase-random states, we mean the ensemble $\Upsilon_{\rm phase} = \{ \ket{\phi} \}_{\mathrm{d}\varphi}$, where the phases $\varphi_n$ are distributed according to the normalized Lebesgue measure given by
\begin{equation}
{\rm d}\varphi =\frac{ {\rm d} \varphi_1}{2 \pi} \cdots \frac{{\rm d}\varphi_{2^N}}{2 \pi}, \notag
\end{equation}
on $[0,2\pi]^{2^N}$.  
This ensemble clearly depends on the choice of amplitudes and basis, which we write $\Upsilon_{\rm phase}({ \{ r_n ,  \ket{u_n} \}_n })$ when there is need to be explicit. 
Note that the ensemble of phase-random ensembles with appropriately distributed amplitudes is the ensemble of random states.

We first point out that studies of thermalization in closed systems \cite{LPSW2009, GLTZ2010} are special instances of the study of phase-random states. 
To see this, consider a Hilbert space $\mathcal{H}$ with dimension $d$, and a Hamiltonian $H=\sum_{n=1}^{d}  e_n \ketbra{ e_n}{ e_n}$. 
The state at time $t$ is given by
\begin{equation}
\ket{\phi(t)} = \sum_n r_n e^{-i { e_n} t / \hbar+ i \varphi_n} \ket{e_n}, \notag
\end{equation}
where $r_n e^{i \varphi_n}=\braket{ e_n}{\phi_0}$
with $r_n \geq 0$ and $\ket{\phi_0}$ is an initial state. 
Then, a time averaged thermodynamical quantity is often considered by assuming {\it phase ergodicity} in the sense that the distribution of phases $e^{-i {e_n} t / \hbar+ i \varphi_n}$ are uniform in $[0, 2\pi]$ in the long-time limit. 
Due to this identification, all studies addressing the time average are equivalent to investigations of statistical properties of the corresponding phase-random states $\Upsilon_{\rm phase}(\{ r_n ,  \ket{e_n} \}_n )$.

For example, in~\cite{LPSW2009} it was proven that time evolution typically gives rise to canonical distributions in subsystems. 
In this case, we consider a Hilbert space $\mathcal{H}_S \otimes \mathcal{H}_E$ where $\mathcal{H}_S$ ($\mathcal{H}_E$) represents a system (environment) with dimension $d_S$ ($d_E$), and $\mathcal{H}_R$ is a restricted subspace constrained by the energy defined by
\begin{equation}
\mathcal{H}_R= \mathrm{span} \{ \ket{ e_{\alpha} }| e-\delta e< e_{\alpha} < e+\delta e \} \subset \mathcal{H}_S \otimes \mathcal{H}_E. \notag
\end{equation}
Then, if an initial state lies in $\mathcal{H}_R$, the reduced density matrix on the system $\tr_{E} \ketbra{\phi(t)}{\phi (t)}$ should be in a neighborhood of $\hat{\rho}_S$ given by
\begin{equation}
\hat{\rho}_S=\sum_{\alpha=1}^{d_R} r_{\alpha}^2 \tr_{E} \ketbra{e_{\alpha}}{e_{\alpha}}, \notag
\end{equation}
for most of the time $t$, where $r_{\alpha}=\mathrm{Re}\braket{e_{\alpha}}{\phi_0}$, $\forall \ket{ e_{\alpha}} \in \mathcal{H}_R$ and $d_R=\mathrm{dim} \mathcal{H}_R$. 
We say that for most of the phase-random states $\Upsilon_{\rm phase}( \{ \tilde{r}_n ,  \ket{e_n} \}_n )$, where $\tilde{r}_n$ is determined by $\ket{\phi(0)}$ and $\mathcal{H}_R$, a reduced density matrix on the system is close to $\hat{\rho}_S$. 

By evaluating the trace distance between the state $\hat{\rho}_S$ and a canonical state $\tr_{E} \mathbb{I}_R/d_R$ where $\mathbb{I}_R$ is the identity matrix on $\mathcal{H}_R$, one can obtain the following condition for $\hat{\rho}_S$ to be a canonical state,
\begin{equation}
\sum_{\alpha,\beta=1}^{d_R}(\tilde{r}_{\alpha}^2 - \frac{1}{d_R})(\tilde{r}_{\beta}^2 - \frac{1}{d_R}) \tr [ \hat{e}^\alpha_S \hat{e}^\beta_S ]=0, \label{Eq:Ther}
\end{equation}
where $\hat{e}^k_S = \tr_{E} \ketbra{e_k}{e_k}$.  This provides a trade-off relation between the initial state and the restricted Hilbert space $\mathcal{H}_R$ for thermalization. In order to see this, consider the following two extreme cases for Eq.~\eqref{Eq:Ther} to hold, which trivially lead to thermalization in subsystems.

Firstly, if the conditions are imposed only on the amplitudes, the amplitudes must be equal, {\it i.e.} $\tilde{r}_{\alpha}=1/d_R$ for all $\alpha$.
Since the amplitudes are defined by the initial state, this is a condition {\it for the initial state} to exhibit thermalization.
On the other hand, if the conditions are imposed only on the restricted Hilbert space $\mathcal{H}_R$, the eigenstates in $\mathcal{H}_R$ should satisfy $\tr [ \hat{e}^\alpha_S \hat{e}^\beta_S ] = \tr [ \hat{e}^{\alpha'}_S \hat{e}^{\beta'}_S]$, $\forall \ket{e_{\alpha}}, \ket{ e_{\beta}},\ket{e_{\alpha'}},\ket{ e_{\beta'}} \in \mathcal{H}_R$. 
By choosing $\alpha = \beta =\beta'$, we obtain
$\tr [ (\hat{e}^\alpha_S)^2 ] = \tr [ \hat{e}^{\alpha}_S \hat{e}^{\alpha'}_S]$. Since the left-hand side is a norm of $\hat{e}^\alpha_S$
and the right-hand side is the Hilbert-Schmidt inner product of $\hat{e}^{\alpha}_S$ and $\hat{e}^{\alpha'}_S$,
it implies that all reduced density matrices should be identical.
Thus, this condition {\it on the eigenstates in $\mathcal{H}_R$} trivially results in thermalization.

Equation~\eqref{Eq:Ther} gives conditions applicable in intermediate situations between these two extreme cases.
It therefore provides grounds for the study of the way thermalization depends on a system's initial state and on its Hamiltonian~\cite{IWU2011}.

\section{Entanglement} \label{Sec:Ent}
We investigate the entanglement properties of phase-random states using entropic measures of entanglement since they can reveal if the state is simulatable by MPSs.  
Consider divisions into two subsystems $A$ and $\bar{A}$, composed of $N_A$ and $N_{\bar{A}}=N-N_A$ qubits respectively. 
We denote the density matrix of $\ket{\phi}$ by $\hat{\phi}=\ketbra{\phi}{\phi}$ and its reduced density matrix on the subsystem $A$ by $\hat{\phi}_A=\tr_{\bar{A}} \hat{\phi} = \tr_{\bar{A}} \ketbra{\phi}{\phi}$.  
For a given pure state $\ket{\phi}$ and subsystem $A$, the amount of entanglement in terms of the linear entropy is given by $E^{(A)}_L (\ket{\phi})=S_L(\hat{\phi}_A)$ where $S_L(\hat{\rho})=1-\tr \hat{\rho}^2$, and in terms of the von Neumann entropy is given by $E^{(A)}(\ket{\phi})=S(\hat{\phi}_A)$ where $S(\hat{\rho}) = -\tr \hat{\rho} \log \hat{\rho}$.   We have $0 \leq E_L^{(A)}(\ket{\phi}) \leq 1-2^{-N_A}$. 
The linear entropy gives a lower bound on the von Neumann entropy, $- \log [1-E^{(A)}_L(\ket{\phi}) ] \leq E^{(A)}(\ket{\phi})$.  
Hence, we will use the linear entropy to measure entanglement in this paper unless otherwise specified.  

For the ensemble of random states $\Upsilon_{\rm rand}$ on $N$ qubits, the average amount of entanglement is 
\begin{align}
\langle E_L^{(A)} \rangle_{\Upsilon_{\rm rand}} &= \int \mathrm{d}\psi E_L^{(A)}(\ket{\psi}) \notag \\
&= 1-\frac{2^{N_A}+2^{N_{\bar{A}}}}{2^N+1}, \label{Eq:RandomAverage}
\end{align}
indicating that random states of large systems are nearly maximally entangled on average~\cite{L1978}.  
For an ensemble of phase random states $\Upsilon_{\rm phase}$ and a choice of subsystem $A$, the average amount of entanglement is defined by 
\begin{equation}
\langle E_L^{(A)} \rangle_{\Upsilon_{\rm phase}} = \int \mathrm{d} \varphi E_L^{(A)}(\ket{\phi}), \notag
\end{equation}
which, recall, is a function of $\{ r_n , \ket{u_n} \}$.
Note that in general $\langle E_L^{(A)} \rangle_{\Upsilon_{\rm phase}} \neq E_L^{(A)} (\hat{\Phi})$, where $\hat{\Phi}$ is the density matrix defined by the phase-random ensemble, namely,
\begin{equation}
\hat{\Phi} = \int \mathrm{d} \varphi \hat{\phi} = \sum_n r_n^2 \ketbra{u_n}{u_n}. \notag
\end{equation}

\begin{figure}[tb]
\centering
\includegraphics[width=85mm, clip]{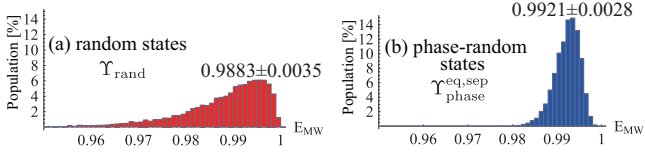} 
  \caption{
(Color online) The distributions for $N=8$ of the amount of entanglement for two ensembles, (a) random states and (b) phase-random states with equal amplitudes and a separable basis, using the Meyer-Wallach measure of entanglement $\EMW(\ket{\phi}):=\frac{2}{N}\sum_{k=1}^N E_L^{(\{k\})}(\ket{\phi})$, where $k$ labels single-qubit subsystems \cite{MW2002}.  The number of samples is $10^4$, binned in intervals of $0.002$.
}
\label{Distribution}
\end{figure}

In order to calculate the average amount of entanglement, we expand the basis elements such that $\ket{u_n}=\sum_{a=1}^{2^{N_A}} \ket{\bar{a}}_A \otimes |\tilde{u}^{(a)}_n \rangle_{\bar{A}}$, where $\{ \ket{\bar{a}}_A \}_{a=1,\cdots, 2^{N_A}}$ is a computational basis for subsystem $A$, $\bar{a}$ is binary for $a-1$, and tildes indicate unnormalized kets.
Defining $\phi^{(A)}_{aa'}$ by
\begin{equation}
\phi^{(A)}_{aa'} := \sum_{n m} r_n r_m e^{i(\varphi_n-\varphi_m)}\braket{\tilde{u}^{(a')}_m}{\tilde{u}^{(a)}_n}, \notag
\end{equation}
$\tr (\hat{\phi}_A)^2$ is given by $\sum_{a,a'} |\phi^{(A)}_{aa'}|^2$.
This is quadratic, so will involve a sum over four basis labels $n,m,l,k$, and where the only phase integral that occurs is 
\begin{equation}
\int \mathrm{d}\varphi e^{i(\varphi_n - \varphi_m + \varphi_k - \varphi_l)} =\delta_{nm}\delta_{kl} + \delta_{nl}\delta_{mk} - \delta_{nm}\delta_{nk}\delta_{nl}. \notag
\end{equation}
Thus we have
\begin{multline}
\langle E_L^{(A)} \rangle_{\Upsilon_{\rm phase}} = \sum_{a,b} \biggl[
\sum_{n,m} r_n^2 r_m^2
(\braket{\tilde{u}^{(b)}_n}{\tilde{u}^{(a)}_n}\braket{\tilde{u}^{(a)}_m}{\tilde{u}^{(b)}_m}\\
+|\braket{\tilde{u}^{(b)}_m}{\tilde{u}^{(a)}_n}|^2) 
-\sum_{l} r_l^4 |\braket{\tilde{u}^{(b)}_l}{\tilde{u}^{(a)}_l}  |^2 \biggr]. \label{SLCal}
\end{multline}
Denoting the mutual information of $\hat{\Phi}$ between $A$ and $\bar{A}$ in terms of the linear entropy by $I_L^{(A)}(\hat{\Phi}) = S_L(\mathrm{Tr}_{\bar{A}} \hat{\Phi})+S_L( \mathrm{Tr}_{A} \hat{\Phi})-S_L(\hat{\Phi})$,
Eq.~\eqref{SLCal} reduces to
\begin{multline}
\langle E_L^{(A)} \rangle_{\Upsilon_{\rm phase}} = I_L^{(A)}(\hat{\Phi})-\sum_{n=1}^{2^N} r_n^4 E_L^{(A)} (\ket{u_n}). \label{AVERAGE}
\end{multline}
Equation~\eqref{AVERAGE} simplifies the investigation of the dependence of $\langle E_L^{(A)} \rangle_{\Upsilon_{\rm phase}}$ on the amplitudes and the basis, $\{ r_n , \ket{u_n} \}$.

We consider two cases in particular.  First, we analyze equal-amplitudes ensembles $\Upsilon_{\rm phase}^\mathrm{eq} = \Upsilon_{\rm phase} ( \{ 2^{-N/2} , \ket{u_n} \})$.
In this case, the average amount of entanglement over phases is given by
\begin{multline}
\langle E_L^{(A)} \rangle_{\Upsilon_{\rm phase}^\mathrm{eq}}
= 1 - \frac{ 2^{N_A} + 2^{N_{\bar{A}}} - 1 }{2^{N}}
-\sum_{n=1}^{2^N} \frac{E_L^{(A)} (\ket{u_n})}{2^{2N}}. \label{UniformAmplitudes}
\end{multline}
This shows that $\langle E_L^{(A)} \rangle_{\Upsilon_{\rm phase}^\mathrm{eq}}$ is a decreasing function of the basis entanglement, $E_L^{(A)}(\ket{u_n})$.
Hence ensembles that also have a separable basis $\{ \ket{u_n^{\rm sep}} \}$, denoted by $\Upsilon_{\rm phase}^{{\rm eq},{\rm sep}} =  \Upsilon_{\rm phase}( {\{ 2^{-N/2} , \ket{u_n^{\rm sep}} \} })$, give the maximum, 
\begin{equation}
\langle E_L^{(A)} \rangle_{\Upsilon_{\rm phase}^{{\rm eq},{\rm sep}}}=1-\frac{{2^{N_A}+2^{N_{\bar{A}}}-1}}{2^{N}}. \notag
\end{equation}
This value is greater than that of random states given by Eq.~\eqref{Eq:RandomAverage}, see also Fig.~\ref{Distribution}. 
For $\ket{\phi^{\rm eq, sep}} = 2^{-N/2} \sum_n e^{i \varphi_n} \ket{u_n^{\rm sep}}$, applying the concentration of measure to $\Delta E_L^{(A)} (\ket{\phi^{\rm eq, sep}}) =|E_L^{(A)} (\ket{\phi^{\rm eq, sep}})- \langle E_L^{(A)} \rangle_{\Upsilon_{\rm phase}^{{\rm eq},{\rm sep}}}|$,
we find
\begin{equation}
\mathrm{Prob} \biggl[\Delta E_L^{(A)} (\ket{\phi^{\rm eq, sep}})> 2/2^{N}+\epsilon \biggr] \leq e^{- c \epsilon^4 2^N}, \label{eq:Conc}
\end{equation}
where $c=1/(2^{11} \pi^2)$.
The proof is similar to that in Ref.~\cite{LPSW2009} (see Appendix \ref{App:Concent} for details).  
Thus the entanglement of phase-random states $\Upsilon_{\rm phase}^{{\rm eq},{\rm sep}}$ is highly concentrated around the average, demonstrated in Fig.~\ref{Distribution}.

States with equal amplitudes in a separable basis are also known as LME states \cite{KK2009}.
These are the class of multipartite states that are maximally entanglable with local auxiliary systems by only local operations.
In Ref. \cite{KK2009}, it is mentioned that the LME states should exhibit high entanglement.  
Our result proves this statement is true in the sense that the uniform ensemble of LME states achieves a higher average amount of entanglement, in terms of linear entropy, than that of random states.

On the other hand, for separable-basis ensembles defined by $\Upsilon_{\rm phase}^{\rm sep} =  \Upsilon_{\rm phase} ( \{ r_n, \ket{u_n^{\rm sep}} \} )$, an upper bound of the average amount of entanglement is given by 
\begin{equation}
0\leq \langle E_L^{(A)} \rangle_{\Upsilon_{\rm phase}^{\rm sep}} \leq S_L(\hat{\Phi}^{\mathrm{sep}}), \notag
\end{equation}
where $\hat{\Phi}^{\mathrm{sep}}=\sum_{n} r_n^2 \ketbra{u_n^{\rm sep}}{u_n^{\rm sep}}$.
When the number of non-zero $r_n$ is $R$, $S_L(\hat{\Phi}^{\mathrm{sep}})$ is bounded by $1-1/R$ from above.  
If $R$ is small, (for instance, if $R=\mathrm{poly}(N)$), the average amount of entanglement cannot be as large as that of random states.  
It is therefore necessary for the basis to be entangled in order to generate a large amount of entanglement on average when $R$ is small.

\section{Simulatability of Hamiltonian dynamics} \label{Sec:Sim}
We now interpret our results in the context of {\it time-independent} Hamiltonian dynamics and consider the simulatability of the state during the time evolution by assuming phase ergodicity.
We consider the area law of entanglement, which states that the von Neumann entropy of entanglement of a large subsystem is at most proportional to its boundary.
Since the breakdown of the area law indicates that the states cannot be simulated by MPSs with a constant matrix size \cite{ECP2010}, the area law gives insight into the simulatability of the state. 
The area law is often studied for ground states of spin systems. 
It is also known that, initial states that do not violate the area law will not do so over a certain time scale evolving under a local Hamiltonian~\cite{BHV2006}.

Applying our results to a lattice of qubits, we consider the long-time average of the von Neumann entropy of entanglement generated by a time-independent Hamiltonian dynamics $\langle E^{(A)} \rangle_{T;\infty}$.
Using the facts that the von Neumann entropy is lower bounded by the linear entropy and the concavity of the logarithm, we obtain a lower bound on $ \langle  E^{(A)} \rangle_{\Upsilon_{\rm phase}}$, which can be identified with $\langle E^{(A)} \rangle_{T;\infty}$ under phase ergodicity. 
Thus, we have $\langle E^{(A)} \rangle_{T;\infty} \geq -\log [ 1- \langle E_L^{(A)} \rangle_{\Upsilon_{\rm phase}}]$.
By applying Eq.~\eqref{AVERAGE}, we can check the area law in the long-time average.

In particular, we consider Hamiltonians composed of separable eigenstates, which are often referred to as {\it semi-classical}.
When the initial state is a superposition of separable eigenstates with equal amplitudes, the initial state is also separable, and the corresponding phase-random states are $\Upsilon_{\rm phase}^{\rm eq, sep}$ which obtains the maximum of Eq.~\eqref{UniformAmplitudes}.
Thus we obtain
\begin{equation}
\langle E^{(A)} \rangle_{T;\infty} \geq N_A -\log(1+2^{2 N_A -N}-2^{N_A -N}), \notag
\end{equation}
which grows in proportion to the volume $N_A$ of the subsystem $A$ when $N_A \ll N$, and not with any boundary size, and the area law is broken.
Since entanglement concentrates around its average during the time evolution as in Eq.~\eqref{eq:Conc}, the states are not simulatable by MPSs with a constant matrix size for most times. 
This is surprising at first because all eigenstates as well as the initial state are separable, however the dynamics generate extremely high entanglement and, thus, is difficult to simulate.

In Ref.~\cite{BHV2006}, timescales necessary for breaking the area law by time evolutions with local Hamiltonians have been studied.
Combined with our result, we can explicitly estimate the timescale necessary for satisfying phase ergodicity when the Hamiltonian is composed of separable eigenstates.

\section{Phase-random circuit}  \label{Sec:PRC}
We present a {\it phase-random circuit} generating an ensemble of states $\Upsilon^{\rm pseudo}_{\rm phase}$ that provides the same average entanglement as the phase-random ensemble $\Upsilon_{\rm phase}^{\rm comp} =  \Upsilon_{\rm phase}( {  \{ r_a ,  \ket{\bar{a}}  \}_a} )$, where $\ket{\phi_0}=\sum r_a  e^{i \varphi_a} \ket{\bar{a}}$ is the input to the circuit and $\{ \ket{\bar{a}} \}_{a=1}^{2^N}$ is the computational basis.

\begin{figure}[tb]
\centering
\includegraphics[width=20em, clip]{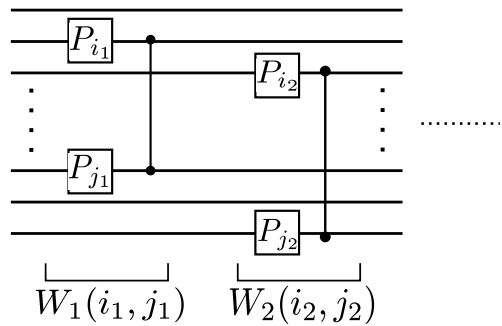}
\caption{Phase-random circuit composed of two-qubit untiaries $W_t (i_t,j_t)$ acting on randomly selected pairs of qubits $(i_t, j_t)$.}
\label{PhaseRandomCircuit}
\end{figure}

A phase-random circuit is similar to those considered in \cite{ODP2007, DOP2007}.  
We consider a circuit composed of $T$ iterations of two-qubit unitaries diagonal in the computational basis denoted by $W_t$, where the subscript $t$ denotes the $t$-th iteration ($t=1, 2, \dots, T$).  For each iteration $t$, the two-qubit unitary $W_t$ acts on a pair of qubits $i,j (j \neq i)$ randomly chosen uniformly from $\{ 1, 2, \cdots , N\}$, and is written
\begin{equation} \label{Wunitary}
W_t = CZ_{i_t j_t} P_{i_t}(\alpha_t) P_{ j_t}(\beta_t),
\end{equation}
where $CZ_{ij} =\mathrm{diag}(1,1,1,-1)$ is a controlled-$Z$ operation on qubits $i$ and $j$, 
$P_k(\theta)=\mathrm{diag}(1,e^{i\theta})$ denotes a phase gate on the qubit $k$, and the
two angles $\alpha, \beta$ are randomly chosen uniformly from the interval $[0,2\pi]$.  
A specific instance of the circuit is described by the set $\{i_t,j_t,\alpha_t,\beta_t\}_{t=1}^T$, and the corresponding output state after $T$ iterations of $W_t$ is given by $\ket{\phi_T} = W_T W_{T-1} \cdots W_{1} \ket{\phi_0}$,  where $\ket{\phi_0}= \sum r_a e^{i \varphi_a} \ket{\bar{a}}$ is an input state in the computational basis $\{\ket{\bar{a}}\}$, defining the ensemble $\Upsilon_{\rm phase}^{\rm comp}$.

\subsection{Summary of results}
Here we state the main results, with the details of the proof to follow.
Denote by $\expectation{E_L^{(A)}(\ket{\phi_T})}$ the expectation value of $ E_L^{(A)}(\ket{\phi_T})$
taken over the uniform distribution of $\{i_t,j_t,\alpha_t,\beta_t\}_{t=1}^T$.
We will prove the following two theorems regarding the ability of $\expectation{ {E}_L^{(A)}(\ket{\phi_T}) }$ to equal the average of phase random states $\langle E_L^{(A)} \rangle_{\Upsilon_{\rm phase}^{\rm comp}}$ after sufficiently many iterations, and about the required number of iterations.

\begin{Theorem}\label{ThmLim}
With the preceding definitions and notations,
\begin{equation}
\lim_{T\rightarrow \infty}\expectation{S_L^{(A)}(\ket{\phi_T})}= \langle E_L^{(A)} \rangle_{\Upsilon_{\rm phase}^{\rm comp}}, \label{MarkovLim}
\end{equation}
with
\begin{multline}
 \langle E_L^{(A)} \rangle_{\Upsilon_{\rm phase}^{\rm comp}}
 =1- \sum_{a,b} r_a^2 r_b^2( \prod_{i \in A} \delta_{a_i b_i} + \prod_{i \in \bar{A}} \delta_{a_i b_i})\\
 + \sum_a r_a^4, \label{AverageLinearEntropy}
\end{multline}
where $a_1 a_2 \cdots a_N \: (b_1 b_2 \cdots b_N) \in \{0,1\}^N$ is a binary representation of $a-1$ ($b-1$). \nonumber 
\end{Theorem}

\begin{Theorem} \label{ThmTime}
Let $T_{\mathrm{mix}}(\epsilon)$ be the number of iterations required to achieve Eq.~\eqref{MarkovLim} with error $\epsilon$, 
namely, 
\begin{equation}
\forall T> T_{\mathrm{mix}}(\epsilon), \biggl| \expectation{S_L^{(A)}(\ket{\phi_T})} -  \langle E_L^{(A)} \rangle_{\Upsilon_{\rm phase}^{\rm comp}} \biggr| < \epsilon. \nonumber 
\end{equation}
For $\Gamma \subset \{1, \cdots, N \}$, define $\kappa^{(\Gamma)}(\ket{\phi_0})$ such that
\begin{equation}
\kappa^{(\Gamma)}(\ket{\phi_0}):=\sum_{a \neq b} r_a^2 r_b^2\prod_{i \in \Gamma}(1-\delta_{a_i b_i}) \prod_{i \notin \Gamma} \delta_{a_i b_i}. \nonumber 
\end{equation}
Then, if $\max_{\Gamma} \kappa^{(\Gamma)}(\ket{\phi_0})=O(2^{-N})$, $T_{\mathrm{mix}}(\epsilon)$ is polynomial in the system size $N$ for any $A$.
In particular, for $r_a \sim p_a(N) 2^{-N/2}$ where $\{ p_a(N) \}_a$ are polynomial functions of $N$,
$T_{\mathrm{mix}}(\epsilon)$ is $poly(N)$.
\end{Theorem}

These results are especially interesting if we consider an ensemble $\Upsilon^{\rm pseudo}_{\rm phase}$ simulating the average amount of entanglement of $\Upsilon_{\rm phase}^{\rm eq, sep}$. 
Since the average entanglement of $\Upsilon_{\rm phase}^{\rm eq, sep}$ violates the area law, most states in $\Upsilon^{\rm pseudo}_{\rm phase}$ do also. Hence, $\Upsilon^{\rm pseudo}_{\rm phase}$ are not simulatable by MPSs although they are generated by a quantum circuit with a polynomial number of elementary gates.

Here, we have focused on the generation of the average amount of entanglement of phase-random states. 
In a separate paper \cite{NMPre}, it is shown that phase-random circuits can approximately generate an ensemble simulating the states themselves.

In the following, we prove Theorems \ref{ThmLim} and \ref{ThmTime} by adapting the method developed in \cite{ODP2007, DOP2007} to the phase-random case.  
In this method, the key technique is to map the evolution of the states in the phase-random circuit to a Markov chain, and so we first briefly review Markov processes in Subsection \ref{IntroMarkov}.
In Subsection \ref{stepA}, we present the map to a Markov chain, and then investigate its stationary distribution.
As, contrary to \cite{ODP2007, DOP2007}, the Markov chain is not irreducible in our case, we first decompose it into irreducible Markov chains in Subsection \ref{stepB}. 
In order to calculate the average amount of entanglement, it is sufficient to consider reduced Markov chains, which are presented in Subsection \ref{stepC}.
By investigating the stationary distribution of the reduced Markov chain, we finally obtain the average 
amount of entanglement after $T$ steps in Subsection \ref{StepNew}.
The mixing time $T_{\mathrm{mix}}(\epsilon)$ for achieving Eq.~\eqref{MarkovLim} is treated in Subsection \ref{stepE}.

\subsection{Introduction of a Markov chain} \label{IntroMarkov}

A Markov chain is a sequence of random variables that take values in a set of {\it states} $S=\{ s \}$, indexed in our case by discrete {\it steps} $t$.  The Markov property is that the probability of $s_{t+1}$ occurring depends only on $s_t$, and is independent of previous states.  We can define at any step $t$ a probability distribution $\Pi_t$ over the states space $S$.  The Markov property then ensures that subsequent distributions are related only to the previous distribution, and that this dependence can be given in the form of a step-independent, stochastic {\it transition matrix} $\mathcal{P}$, with matrix elements denoted by $\mathcal{P}(s,s')$.  Thus, the probability distribution at step $t$ is given by $\Pi_t = \mathcal{P}^t \Pi_0$, where $\Pi_0$ is an initial distribution.

When a Markov chain is {\it irreducible} and {\it aperiodic}, the probability distribution on each state
converges after sufficiently many steps.  That is, for all $s$, there exists a unique $\Pi_\infty(s)=\lim_{t \rightarrow \infty} \Pi_t(s)$ that is independent of the initial probability distribution. 
Irreducibility is a property of the transition matrix implying that 
any state $s$ can transition to any other state in a finite number of steps, that is,
for all $s$ and $s'$, there exists a $t$ such that $\mathcal{P}^t(s,s')>0$.
Aperiodicity implies that, for all states $s$, there exists a non-zero probability to remain in that state, namely, $\mathcal{P}(s,s)>0$ for all $s$.
A sufficient condition for a distribution to be stationary is given by the {\it detailed balance} condition
\begin{equation}
\Pi(s) \mathcal{P}(s,s')=\Pi(s') \mathcal{P}(s',s), \text{ for all } s,s' \in S. \nonumber 
\end{equation}
When a Markov chain satisfies the detailed balance equations, it is referred to as {\it reversible}.

Next, we define the {\it mixing time}, which is the number of Markov chain steps required for the distance between the actual distribution and the stationary distribution to be small, where we define the distance between two probability distributions as follows.
Let $\delta(s_0)$ be an initial probability distribution of a Markov chain with value $1$ at $s_0$ and zero elsewhere on the state space $S$.  
Let us denote the sum of the probabilities of a distribution over a subset of states $S'$ by $\Pi(S')=\sum_{s \in S'} \Pi(s)$, and by $\Pi_t(S' |\delta(s_0))$ such a sum at step $t$ of a Markov chain that initialized with the distribution $\delta(s_0)$.
The {\it variation distance} after $t$-steps is defined by
\begin{equation}
\Delta_{s_0}(t):=\max_{S' \subseteq S} |\Pi_t(S' |\delta(s_0) ) - \Pi_\infty(S')|. \nonumber 
\end{equation}
The mixing time $T_{\mathrm{mix}}(\epsilon)$ is then defined for any $\epsilon >0$ by 
\begin{equation}
T_{\mathrm{mix}}(\epsilon) := \min\{t | \max_{s_0 \in S} \Delta_{s_0}(t') \leq \epsilon \text{ \ for all \ } t' \geq t \}. \nonumber 
\end{equation}
This is the number of steps it would take to get $\epsilon$-close to the stationary distribution in the worst case.  In practice, we do not actually use this definition of the mixing time, but rather the following Theorem~\ref{MixingTime} and Corollary~\ref{AbsoluteGap} regarding the transition matrix.

For a transition matrix $\mathcal{P}$ of a reversible Markov chain, let us label the eigenvalues of $\mathcal{P}$ in decreasing order such that
\begin{equation}
1=\lambda_1 > \lambda_2 > \cdots. \nonumber 
\end{equation}
Then, $\eta :=1-\lambda_2$ is called its {\it absolute spectral gap}. The absolute spectral gap $\eta$ gives 
an upper bound on the mixing time as stated in the following theorem.
\begin{Theorem}[Theorem 12.3 in \cite{LPW2008}] \label{MixingTime}
Let $\mathcal{P}$ be the transition matrix of a reversible Markov chain on $S$, and let $\Pi(\min):=\min_{s \in S} \Pi(s)$.
Then
\begin{equation}
T_{\mathrm{mix}}(\epsilon) \leq \log(\frac{1}{\epsilon \Pi(\min)}) \frac{1}{\eta}. \nonumber 
\end{equation}
\end{Theorem}

Moreover, a lower bound on the absolute spectral gap $\eta$ is obtained by the {\it canonical path} method.
Viewing a reversible transition matrix $\mathcal{P}$ as a graph with vertex set $S$, define the edge set $E=\{(s,s') | \mathcal{P}(s,s')>0 \}$. 
A canonical path from $s$ to $s'$ is a sequence $\mathcal{E}_{ss'} =(e_1,\cdots,e_m)$ of edges in $E$ such that $e_1=(s,s_1)$, $e_2=(s_1,s_2)$, $\cdots$, $e_m=(s_{m-1},s')$ for vertices $s_i$, $i=1,2,\cdots,m$.  We have the following Corollary~\ref{AbsoluteGap}.
\begin{Corollary}[Corollary 4 in \cite{S1992}] \label{AbsoluteGap}
For a given transition matrix $\mathcal{P}$, let $Q(s,s'):= \Pi_\infty(s) \mathcal{P}(s,s')$ and 
\begin{equation}
\rho:= \max_{e \in E} \frac{1}{Q(e)}\sum_{\begin{subarray}{c} s,s' \\ \mathcal{E}_{ss'} \ni e \end{subarray}} \Pi_\infty(s) \Pi_\infty(s'). \label{InverseGap} 
\end{equation}
Then
\begin{equation}
\frac{1}{8\rho^2} \leq \eta. \nonumber 
\end{equation}  
\end{Corollary}

By combining Theorem \ref{MixingTime} and Corollary \ref{AbsoluteGap}, an upper bound on the mixing time can be obtained.

\subsection{Map to a Markov chain} \label{stepA}

We will now show that the change in the state $\ket{\phi_t} \rightarrow \ket{\phi_{t+1}}$ upon the application of the two-qubit unitary $W_{t+1}$ defined by Eq.~(\ref{Wunitary}) can be formulated in terms of a transition matrix action on the indices of expansion coefficients of the state in the basis of local Pauli operators.  The hermiticity of this basis ensures that the coefficients are real, and hence their square gives a valid probability distribution, while its locality ensures that we can focus on the qubits $i$ and $j$ where $W_{t+1}$ acts, eventually simplifying the calculation of the linear entropy.

Let us consider the expansion of $\ketbra{\phi_t}{\phi_t}$ given by
\begin{equation}
\ketbra{\phi_t}{\phi_t} = \frac{1}{2^{N/2}} \sum_{q_1, \cdots, q_N} \xi_t(q_1, \cdots, q_N) 
\sigma_{q_1} \otimes \cdots \otimes \sigma_{q_N}, \nonumber 
\end{equation} 
where $q_i \in \{ 0, x, y, z \}$ and $\sigma_{q_i}$ are Pauli operators.
We denote $(q_1, \cdots, q_N)$ by the vector $\mathbf{q}$. 
We construct a Markov chain defined on $\{ \mathbf{q} \}$ in which the probability distribution
is given by the expectation value of $\xi_t^2(\mathbf{q})$ over $\alpha_t$ and $\beta_t$, which is denoted by $\expectation{\xi_t^2(\mathbf{q})}$. 
For this purpose, we first examine $\expectation{\xi_t^2(\mathbf{q})}$, and then construct the Markov chain.
For simplicity, hereafter we omit the step indices on qubits and write $(i, j)$.

By applying $W_{t+1}$ on a randomly chosen pair of qubits $(i, j)$,
the coefficients $\{ \xi_{t+1}(\mathbf{q}) \}$ of the state $\ketbra{\phi_{t+1}}{\phi_{t+1}}$ become 
\begin{align}
\xi_{t+1}(\mathbf{p}) &= \frac{1}{4} \sum_{q_{i},q_{j}}
\xi_t(\mathbf{p}_{p_i \rightarrow q_i, p_j \rightarrow q_j}) \times \nonumber  \\
&\tr[\sigma_{p_{i}} \otimes \sigma_{p_{j}} W_{t+1} \sigma_{q_{i}} \otimes \sigma_{q_{j}}
W_{t+1}^{\dagger}], \nonumber 
\end{align}
where $\mathbf{p}_{p_i \rightarrow q_i, p_j \rightarrow q_j}$ is $\mathbf{p}$ but with components $(p_{i}, p_{j})$ replaced by $(q_{i}, q_{j})$.
Squaring this to arrive at a probability distribution, we have
\begin{align}
\xi_{t+1}^2(\mathbf{p}) = \frac{1}{4} \sum_{\begin{subarray}{c} q_{i},q_{j},\\q'_{i},q'_{j}\end{subarray}}
\xi_t(&\mathbf{p}_{p_i \rightarrow q_i, p_j \rightarrow q_j})  \notag \\ 
& \times\xi_t(\mathbf{p}_{p_i \rightarrow q'_i, p_j \rightarrow q'_j}) G_{t+1}(\mathbf{p},\mathbf{q},\mathbf{q'}), \label{D5}
\end{align}
where
\begin{multline}
G_{t+1}(\mathbf{p},\mathbf{q},\mathbf{q'}):= \\
\tr[\sigma_{p_{i}} \otimes \sigma_{p_{j}} W_{t+1} \sigma_{q_{i}} \otimes \sigma_{q_{j}}
W_{t+1}^{\dagger}]\\
\times \tr[\sigma_{p_{i}} \otimes \sigma_{p_{j}} W_{t+1} \sigma_{q'_i} \otimes \sigma_{q'_j}
W_{t+1}^{\dagger}]. \nonumber 
\end{multline}

In order to see that $G_{t+1}$ defines a transition matrix, it is important to recognize that it treats the sets of Pauli indices $\{0,z\}$ and $\{x,y\}$ equivalently.  We write $w_{0z}$ and $w_{xy}$ for arbitrary elements of each set respectively, and we define an involution $\neg$ as $\neg 0=z$ and $\neg x=y$.
Averaging over $\alpha_t, \beta_t$, we obtain
\begin{equation}
\expectation{G_{t+1}(\mathbf{p},\mathbf{q},\mathbf{q'})}=\delta_{\mathbf{q} \mathbf{q'}} \times
\begin{cases}
16 & \text{case I} \\
8 & \text{case II} \\
4 & \text{case III} \\
0 & \text{otherwise},
\end{cases}
\end{equation}
where each case is defined by
\begin{align}
&\text{case I} \Leftrightarrow (p_i=q_i=w_{0z}) \wedge (p_j=q_j=w_{0z}) \nonumber  \\
&\text{case II} \Leftrightarrow (\neg p_i=q_i=w_{0z}) \wedge (p_j,q_j=w_{xy}) \notag \\
&\ \ \ \ \ \ \ \ \ \ \ \ \ \ \ \lor \ \  (p_i,q_i=w_{xy}) \wedge (\neg p_j=q_j=w_{0z})  \nonumber  \\
&\text{case III} \Leftrightarrow p_i, q_i, p_j, q_j=w_{xy}. \nonumber 
\end{align}

By substituting $\expectation{G_{t+1}(\mathbf{p},\mathbf{q},\mathbf{q'})}$ into Eq.~\eqref{D5},
we obtain $\expectation{\xi_{t+1}^2(\mathbf{p}) | \ket{\phi_t}}$, the expectation
value conditional on state $\ket{\phi_t}$, with values as shown in Table~\ref{Exp} where
\begin{align}
&A_t=\xi_t(\mathbf{p}),\nonumber  \\
&B_t(w)=\frac{1}{2}\sum_{w'=x,y} \xi_t(\mathbf{p}_{p_i \rightarrow \neg w, p_j \rightarrow w'}),\nonumber \\
&C_t(w)=\frac{1}{2}\sum_{w'=x,y} \xi_t(\mathbf{p}_{p_i \rightarrow w', p_j \rightarrow \neg w}),\nonumber  \\
&D_t=\frac{1}{4}\sum_{w=x,y} \sum_{w'=x,y} \xi_t(\mathbf{p}_{p_i \rightarrow w, p_j \rightarrow w'}) \nonumber . 
\end{align}

\begin{table}
\begin{tabular}{c|c|cccc}
\multicolumn{2}{c|}{} & \multicolumn{4}{c}{$p_i$} \\ \cline{3-6}
\multicolumn{2}{c|}{%
\raisebox{.5\normalbaselineskip}[0pt][0pt]{ $\expectation{\xi_{t+1}^2$}}} & 0 & x & y & z \\ \hline
 & 0 & $A_t$ & $C_t(0)$ & $C_t(0)$ & $A_t$ \\ 
 & x & $B_t(0)$ & $D_t$ & $D_t$ & $B_t(z)$ \\  
 \raisebox{.5\normalbaselineskip}[0pt][0pt]{$p_j$} & y & $B_t(0)$ & $D_t$ & $D_t$ & $B_t(z)$ \\ 
 & z &$A_t$ & $C_t(z)$ & $C_t(z)$ & $A_t$ \\ 
\end{tabular} 
\caption{Table of $\expectation{\xi_{t+1}^2(\mathbf{p})| \ket{\phi_t}}$ as a function of $p_i$ and $p_j$.} 
\label{Exp}
\end{table}

We are now prepared to define a Markov chain:
\begin{Definition}[Markov chain $\mathcal{M}$]
Let $\mathcal{M}$ be a Markov chain on a set $S=\{ 0, x, y, z\}^N = \{ \mathbf{q} \}$.
The transition process is described as follows.
In each step, $i$ and $j$ are randomly chosen from $\{ 1, \cdots , N\}$ and
the transition from $\mathbf{q} \in S$ to $\mathbf{p} \in S$ occurs probabilistically according to Table~\ref{tab:transprob}.
\begin{table}[h!]
\begin{tabular}{c|c|c}
$(q_i, q_j)$ & $(p_i,p_j)$ & Probability  \\
\hline
\hline
$(w_{0z},w_{0z})$ & $(q_i,q_j)$ & 1 \\
\hline
$(w_{0z},w_{xy})$ & $(\neg q_i,x)$ & 1/2 \\
				& $(\neg q_i,y)$ & 1/2 \\
\hline
$(w_{xy},w_{0z})$ & $(x,\neg q_j)$ & 1/2 \\
				& $(y,\neg q_j)$ & 1/2 \\
\hline
$(w_{xy},w_{xy})$ & $(x,x)$ & 1/4 \\
				& $(x,y)$ & 1/4 \\
				& $(y,x)$ & 1/4 \\
				& $(y,y)$ & 1/4
\end{tabular}
\caption{Transition probabilities.}
\label{tab:transprob}
\end{table}
The transition probability from $\mathbf{q}$ to $\mathbf{p}$ and the probability distribution over $\mathbf{p}$ after $t$ steps are denoted by $\mathcal{P}(\mathbf{q},\mathbf{p})$ and $\Pi_t(\mathbf{p})$, respectively.
The initial distribution $\Pi_0(\mathbf{p})$ is identified with $\xi_0^2(\mathbf{p})$.
\end{Definition}

\begin{Proposition}
The probability distribution $\Pi_t(\mathbf{p})$ of the Markov chain $\mathcal{M}$ coincides with
$\expectation{\xi_{t}^2(\mathbf{p})| \ket{\phi_0}}$.
\end{Proposition}

Since the initial distribution of the Markov chain $\mathcal{M}$ is given by $\xi_0^2(\mathbf{p})$,
\begin{align}
\Pi_1(\mathbf{p})&=\sum_{\mathbf{r}}\mathcal{P}(\mathbf{p},\mathbf{r}) \Pi_0(\mathbf{r}) \nonumber  \\
&=\sum_{\mathbf{r}}\mathcal{P}(\mathbf{p},\mathbf{r}) \xi_{0}^2(\mathbf{r}) \nonumber  \\
&=\expectation{  \xi_1^2(\mathbf{p}) | \ket{\phi_{0}}}, \nonumber 
\end{align}
where the last equation is obtained using Table~\ref{Exp} with the definition of the Markov chain $\mathcal{M}$.
By induction on $t$, Proposition 1 is proven.  For example
\begin{align}
\Pi_2(\mathbf{p})&=\sum_{\mathbf{r}}\mathcal{P}(\mathbf{p},\mathbf{r}) \Pi_1(\mathbf{r}) \nonumber  \\
&=\sum_{\mathbf{r}}\mathcal{P}(\mathbf{p},\mathbf{r}) \expectation{  \xi_1^2(\mathbf{r}) | \ket{\phi_{0}}} \nonumber  \\
&=\expectation{\sum_{\mathbf{r}}\mathcal{P}(\mathbf{p},\mathbf{r}) \xi_1^2(\mathbf{r}) | \ket{\phi_{0}}}\nonumber  \\
&=\expectation{\xi_2^2(\mathbf{p}) | \ket{\phi_{0}}}. \nonumber 
\end{align}

We recall that a probability distribution $\Pi$ can be viewed as a vector in a $4^N$-dimensional space, which we'll call $V_S$, where $S=\{0,x,y,z\}^N$.  For a given $t$, the set of all possible $\Pi_t$ comprise the probability simplex in $V_S$ defined by $\sum_\mathbf{p} \Pi_t(\mathbf{p})=1$.  The transition rules given in Tables~\ref{Exp},\ref{tab:transprob} define a transition matrix $\mathcal{P}$ on $V_S$ with matrix elements written as $\mathcal{P}(\mathbf{q},\mathbf{p})$.

\subsection{Irreducible decomposition of the Markov chain} \label{stepB}

\begin{figure}[tb]
   \centering
   \includegraphics[width=25em, clip]{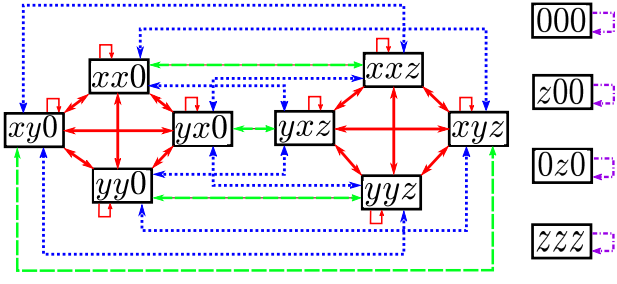}
   \caption{(Color online) An example of some irreducible sets of $\mathcal{M}$ when $N=3$. 
	Directed lines imply the transition occurs with a fixed probability.
	The probability of blue (dotted), red (solid), green (dashed) and purple (dashed-dotted) lines is $1/6$, $1/12$, $1/3$ and $1$, respectively.  
	Elements such as $000$, $z00$ and so on are invariant under the Markov process.}
   \label{MarkovChainFig}
  \end{figure}

In this subsection, we give the irreducible decomposition of $V_S$.
By the definition of the Markov chain $\mathcal{M}$, 
it is obvious that the number of $x$ and $y$ in $\mathbf{q}$ is invariant under the action of the transition matrix $\mathcal{P}$.
Thus we obtain the irreducible decomposition of $V_S$ given by Proposition \ref{PropIr} (see also Fig.~\ref{MarkovChainFig}).

\begin{Proposition}[Irreducible decomposition of $S$]\label{PropIr}
For $\mathbf{q}=q_1q_2 \cdots q_N \in S=\{ 0,x,y,z \}^N$,
let $X(\mathbf{q})$ be the sequence $\{i \in [1, \cdots, N] | q_i \in \{ x,y \}\}$ and
let $S(\Gamma)$ be the set defined by
\begin{equation}
S(\Gamma):=\{\mathbf{q} | X(\mathbf{q})= \Gamma\}, \nonumber 
\end{equation}
where $\Gamma$ is any subset of $\{1,2,\cdots, N\}$.
Then, for the Markov chain $\mathcal{M}$, the irreducible decomposition of $V_S$ is given by
\begin{equation}
V_S=\underset{\mathbf{q} \in S(\emptyset)}{\oplus} V_{\{\mathbf{q} \}} \underset{\Gamma \neq \emptyset}{\oplus} V_{S(\Gamma)}, \nonumber 
\end{equation}
where $V_{S'}$ is the vector space defined by the subset $S'$.
\end{Proposition}

Since $V_{\{\mathbf{q} \}}$ is always one dimensional by definition, we have that $\Pi_t(\mathbf{q})=\Pi_0(\mathbf{q})$ for all $\mathbf{q} \in S(\emptyset)$ and for all $t$.
Thus $\Pi_t(\mathbf{q} \in S(\emptyset))$ is given by
\begin{align}
\Pi_t(\mathbf{q})&=\Pi_0(\mathbf{q}) = \xi_0(\mathbf{q})^2 \nonumber \\
&= 2^{-N} \bra{\phi_0} \sigma_{\mathbf{q}} \ket{\phi_0}^2  \nonumber  \\
&=2^{-N} \sum_{a,b} r_a^2 r_b^2 \prod_{i=1}^N [ \delta_{q_i 0} + \delta_{q_i z}(1-2a_i)(1-2b_i)], \label{simplecase}
\end{align}
where we have used the fact that, for $\mathbf{q} \in S(\emptyset)$, $q_i \in \{0,z \}$ for all $i$ and $\sigma_{\mathbf{q}} := \sigma_{q_1} \otimes \cdots  \otimes \sigma_{q_N}$.

\subsection{Reduction of the Markov chain} \label{stepC}

In order to describe the evolution of $E_L^{(A)} (\ket{\phi_t})$,
a full investigation of the Markov chain $\mathcal{M}$ is not necessary  due to the definition of the linear entropy $S_L (\hat{\rho}) = 1 - \tr \hat{\rho}^2$. 
This can be seen by considering the reduced density matrix of $\ket{\phi_t}$ on a subsystem $A$.
\begin{align}
\hat{\phi}_A^{(t)}&=\tr_{\bar{A}} \ketbra{\phi_t}{\phi_t} \nonumber  \\
 &= \frac{1}{2^{N/2}} \sum_{\mathbf{q}} \xi_t(\mathbf{q}) 
\tr_{\bar{A}} \sigma_\mathbf{q}, \nonumber 
\end{align}
and $\tr (\hat{\phi}_A^{(t)})^2$ is given by
\begin{equation}
\tr (\hat{\phi}_A^{(t)})^2=2^{N_{\bar{A}}} 
\sum_{\begin{subarray}{c} \mathbf{q} \text{ s.t.} \\ q_i=0, i \in \bar{A} \end{subarray}} 
\xi_{t} (\mathbf{q})^2. \nonumber 
\end{equation}
Hence, its expectation value is 
\begin{align}
\mathbf{E}[\tr (\hat{\phi}_A^{(t)})^2 | \ket{\phi_0}]&=
2^{N_{\bar{A}}} 
\sum_{\begin{subarray}{c} \mathbf{q} \text{ s.t.} \\ q_i=0, i \in \bar{A} \end{subarray}} 
\mathbf{E}[\xi_t(\mathbf{q})^2| \ket{\phi_0}] \nonumber  \\
&=2^{N_{\bar{A}}} \sum_{\begin{subarray}{c} \mathbf{q} \text{ s.t.} \\ q_i=0, i \in \bar{A}\end{subarray}} 
\Pi_t(\mathbf{q}). \nonumber 
\end{align}
Thus, it is sufficient to investigate $\Pi_t(\mathbf{q})$ for $\mathbf{q}$ such that 
$q_i=0$ for $i \in \bar{A}$.
The only important property is the number of non-zero terms in 
$\mathbf{q}$.
For this reason, let us define the set $\chi^{(\Gamma)} (\mathbf{q})$ as
\begin{equation}
\chi^{(\Gamma)}(\mathbf{q}):=\{ i \in [1,\cdots, N] | q_i \neq 0, \mathbf{q} \in S(\Gamma) \}, \nonumber 
\end{equation}
which indicates the positions of non-zero terms in $\mathbf{q} \in S(\Gamma)$.
Using this notation, the expectation value is written by 
\begin{equation}
\mathbf{E}[\tr (\rho_A^{(t)})^2 | \ket{\phi_0}]=
2^{N_{\bar{A}}} \sum_{\Gamma \subset A}
\sum_{\begin{subarray}{c} \mathbf{q} \text{ s.t.} \\ \chi^{(\Gamma)}(\mathbf{q})=A \end{subarray}} \Pi_t(\mathbf{q}). \label{Calculation}
\end{equation}
Since $\Pi_t(\mathbf{q})$ for $\mathbf{q} \in S(\emptyset)$ is already given
by Eq.~\eqref{simplecase}, we consider only $\mathbf{q} \in S(\Gamma)$ for $\Gamma \neq \emptyset$.

\begin{figure}[tb]
   \centering
   \includegraphics[width=25em, clip]{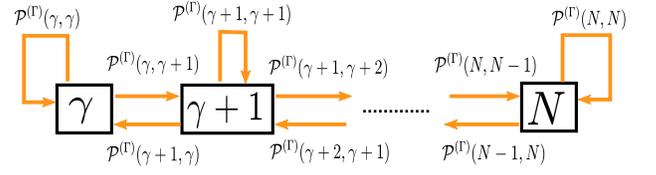}
   \caption{(Color online) Graph of the Markov chain $\mathcal{\tilde{M}}_{\Gamma}$. 
	Directed colored lines imply transition occurs with a fixed probability.
	Transition probabilities are given in Proposition 3.}
   \label{newMarkov}
  \end{figure}

For this reason, we can reduce the Markov chain $\mathcal{M}$ to a simpler Markov chain $\mathcal{\tilde{M}}_\Gamma$.
For a given number of $x$ or $y$ entries $\gamma:=|\Gamma|$ in $\mathbf{q}$, the number of non-zero elements $|\chi^{(\Gamma)}(\mathbf{q})|$ can take values $\{\gamma, \gamma+1, \cdots , N\}$.  The new Markov chain is a drunkard's walk on this set (see Fig,~\ref{newMarkov}), with transition probabilities given by the following proposition.

\begin{Proposition}
For the Markov chain $\mathcal{\tilde{M}}_{\Gamma}$ defined on $\{ i \in \{ \gamma ,\cdots N \} \}$, a transition from $i$ to $j$ occurs with probability,
\begin{align}
&\mathcal{P}^{(\Gamma)}(i, j=i+1) =\frac{2 \gamma (N-i)}{N(N-1)}, \nonumber  \\
&\mathcal{P}^{(\Gamma)}(i, j=i-1) =\frac{2\gamma (i-\gamma)}{N(N-1)}, \nonumber  \\
&\mathcal{P}^{(\Gamma)}(i, j=i) =\frac{ \gamma(\gamma-1)+ (N-\gamma)(N-\gamma-1)}{N(N-1)}. \nonumber 
\end{align}
\end{Proposition}

This is directly induced from the definitions of the Markov chain $\mathcal{M}$ and $\mathcal{\tilde{M}}$.
For instance, the transition $i \rightarrow i+1$ in $\mathcal{\tilde{M}}_{\Gamma}$ occurs if and only if
$(q_i,q_j)=(w_{xy},0)$ or $(q_i,q_j)=(0, w_{xy})$ in $\mathcal{M}$. As
the number of zeroes is $N-i$ and the number of $w_{xy}=\gamma$,
its probability is given by $\frac{\gamma (N-i)}{N(N-1)/2}$.

As stated in the introductory subsection, since the Markov chain $\mathcal{\tilde{M}}_{\Gamma}$ is 
irreducible and aperiodic, it has a unique stationary distribution $\Pi^{(\Gamma)}_\infty$, which is
determined by the detailed balance condition and the normalization.
The detailed balance condition gives the equation
\begin{equation}
\Pi^{(\Gamma)}_\infty(i) \mathcal{P}^{(\Gamma)}(i, i+1) 
=\Pi^{(\Gamma)}_\infty(i+1) \mathcal{P}^{(\Gamma)}(i+1, i). \nonumber 
\end{equation}
Using this equation, we obtain 
\begin{equation}
\Pi^{(\Gamma)}_\infty(i) 
=\begin{pmatrix} N-\gamma \\ i- \gamma \end{pmatrix}\Pi^{(\Gamma)}_\infty(\gamma). \label{A}
\end{equation}
The normalization in $S(\Gamma)$ depends on the input state $\ket{\phi_0}$ as
\begin{equation}
\sum_{i=\gamma}^N \Pi^{(\Gamma)}_\infty(i) = 
\sum_{\mathbf{q} \in S(\Gamma)} \Pi_0(\mathbf{q}) = 
\sum_{\mathbf{q} \in S(\Gamma)} \xi_0^2(\mathbf{q}). \label{B}
\end{equation}
On the other hand, Eq.~\eqref{A} gives
\begin{equation}
\sum_{i=\gamma}^N \Pi^{(\Gamma)}_\infty(i) = 2^{N-\gamma} \Pi^{(\Gamma)}_\infty(\gamma). \nonumber 
\end{equation}
Hence, the stationary distribution is given by
\begin{equation}
\Pi^{(\Gamma)}_\infty(i) 
=\frac{ 1}{2^{N-\gamma} }\begin{pmatrix} N-\gamma \\ i- \gamma \end{pmatrix}\sum_{\mathbf{q} \in S(\Gamma)} \xi_0^2(\mathbf{q}). \nonumber 
\end{equation}

Recalling that $\xi_0^2(\mathbf{q})=\bra{\phi_0} \sigma_{\mathbf{q}} \ket{\phi_0}$, 
it is not difficult to compute $\sum_{\mathbf{q} \in S(\Gamma)} \xi_0^2(\mathbf{q})$, 
which gives
\begin{equation}
\sum_{\mathbf{q} \in S(\Gamma)} \xi_0^2(\mathbf{q}) = \sum_{a \neq b} r_a^2 r_b^2
\prod_{i \in \Gamma}(1-\delta_{a_i b_i}) \prod_{i \notin \Gamma} \delta_{a_i b_i}. \nonumber 
\end{equation}
Thus the stationary distribution for a given subset $\Gamma$ is
\begin{equation}
\Pi^{(\Gamma)}_\infty(i) 
=\frac{ 1}{2^{N-\gamma} }\begin{pmatrix} N-\gamma \\ i- \gamma \end{pmatrix}\sum_{a \neq b} r_a^2 r_b^2
\prod_{i \in \Gamma}(1-\delta_{a_i b_i}) \prod_{i \notin \Gamma} \delta_{a_i b_i}. \label{StationaryDistribution}
\end{equation}

\subsection{Calculation of $\lim_{T \rightarrow \infty} \mathbf{E}[S_L^{(A)} (\ket{\phi_T})]$} \label{StepNew}

We will now calculate the large time limit of the expectation value of the amount of entanglement $\lim_{T \rightarrow \infty} \mathbf{E}[{ E}_L^{(A)} (\ket{\phi_T})]$ using the results of the previous two subsections.
From Eq.~\eqref{Calculation} we have
\begin{align}
&\mathbf{E}[\tr (\hat{\phi}_A^{(T)})^2 | \ket{\phi_0}] \nonumber  \\
&=
2^{N_{\bar{A}}} \sum_{\Gamma \subset A}
\sum_{\begin{subarray}{c} \mathbf{q} \text{ s.t.} \\ \chi^{(\Gamma)}(\mathbf{q}) \subset A \end{subarray}} \Pi_T(\mathbf{q}) \nonumber  \\
&=
2^{N_{\bar{A}}} \biggl[
\sum_{\begin{subarray}{c} \mathbf{q} \text{ s.t.} \\ \chi^{(\emptyset)}(\mathbf{q}) \subset A \end{subarray}} +
\sum_{A \supset \Gamma \neq \emptyset}
\sum_{\begin{subarray}{c} \mathbf{q} \text{ s.t.} \\ \chi^{(\Gamma)}(\mathbf{q}) \subset A \end{subarray}} \biggr]\Pi_T(\mathbf{q}). \label{Goal}
\end{align}

The first term in Eq.~\eqref{Goal} is calculated from Eq.~\eqref{simplecase} as
\begin{align}
&\sum_{\begin{subarray}{c} \mathbf{q} \text{ s.t.} \\ \chi^{(\emptyset)}(\mathbf{q})\subset A \end{subarray}}\Pi_T(\mathbf{q}) \nonumber  \\
&=\sum_{\begin{subarray}{c} \mathbf{q} \text{ s.t.} \\ \chi^{(\emptyset)}(\mathbf{q})\subset A \end{subarray}}
2^{-N} \sum_{a,b} r_a^2 r_b^2 \prod_{i=1}^N [ \delta_{q_i 0} + \delta_{q_i z}(1-2a_i)(1-2b_i)] \nonumber  \\
&=2^{-N_{\bar{A}}} \sum_{a,b}r_a^2 r_b^2 \prod_{i \in A} \delta_{a_i b_i}, \nonumber 
\end{align}
where the last expression is derived from the relation
\begin{align}
&\sum_{\begin{subarray}{c} \mathbf{q} \text{ s.t.} \\ \chi^{(\emptyset)}(\mathbf{q}) \subset A \end{subarray}}
\prod_{i=1}^N [ \delta_{q_i 0} + \delta_{q_i z}(1-2a_i)(1-2b_i)] \nonumber  \\
&=2^{N_{A}} \prod_{i \in A} \delta_{a_i b_i}. \nonumber 
\end{align}

The second term in Eq.~\eqref{Goal} is obtained from the stationary distributions $\Pi^{(\Gamma)}_\infty(i)$ given by Eq.~\eqref{StationaryDistribution}. 
From the definition of the Markov chain $\mathcal{M}$, for $\mathbf{q}, \mathbf{q'} \in S(\Gamma)$, if the number of $z$ in $\mathbf{q}$ is equal to that in $\mathbf{q'}$, $\Pi_\infty(\mathbf{q})=\Pi_\infty(\mathbf{q'})$,
so that
\begin{align}
&\sum_{A \supset \Gamma \neq \emptyset}
\sum_{\begin{subarray}{c} \mathbf{q} \text{ s.t.} \\ \chi^{(\Gamma)}(\mathbf{q})\subset A \end{subarray}}\Pi_\infty(\mathbf{q}) \nonumber \\
&=\sum_{A \supset \Gamma \neq \emptyset} \sum_{i=1}^{N_A} 
\frac{(\begin{subarray}{c} N_A- \gamma \\ i-\gamma  \end{subarray} )}{(\begin{subarray}{c} N- \gamma \\ i-\gamma  \end{subarray})} \Pi^{(\Gamma)}_\infty(i) \nonumber \\
&= \sum_{A \supset \Gamma \neq \emptyset} \sum_{a \neq b} r_a^2 r_b^2 \sum_{i=\gamma}{N_A} 
(\begin{subarray}{c} N_A- \gamma \\ i-\gamma  \end{subarray} ) \prod_{i \notin \Gamma} \delta_{a_i b_i} \prod_{i \in \Gamma} (1-\delta_{a_i b_i}) \nonumber  \\
&=2^{-N_{\bar{A}}} \sum_{a \neq b}r_a^2 r_b^2 \prod_{i \in \bar{A}} \delta_{a_i b_i}, \nonumber 
\end{align}
where we have used the relation
\begin{equation}
\sum_{A \supset \Gamma \neq \emptyset} \prod_{i \notin \Gamma} \delta_{a_i b_i} \prod_{i \in \Gamma} (1-\delta_{a_i b_i})
=-\delta_{ab}+\prod_{i \in \bar{A}} \delta_{a_i b_i}. \nonumber 
\end{equation}

Combining the two we arrive at the final expression
\begin{multline}
\lim_{T\rightarrow \infty} \mathbf{E}[\tr (\hat{\phi}_A^{(T)})^2 | \ket{\phi_0}]\\
=\sum_{a,b}r_a^2 r_b^2 (\prod_{i \in A} \delta_{a_i b_i}+  \prod_{i \in \bar{A}} \delta_{a_i b_i})
-\sum_{a}r_a^4, \nonumber 
\end{multline}
and since ${E}_L^{(A)} (\ket{\phi_T})=1-\tr (\hat{\phi}_A^{(T)})^2 $, Eq.~\eqref{AverageLinearEntropy} is obtained.

\subsection{Mixing time} \label{stepE}

In this final subsection we bound the mixing time of the Markov chain $\mathcal{\tilde{M}}_{\Gamma}$
using Theorem \ref{MixingTime} and Corollary \ref{AbsoluteGap}.

From Eq.~\eqref{StationaryDistribution}, the stationary distribution in $S(\Gamma)$ is given by
\begin{equation}
\Pi^{(\Gamma)}_\infty(i) 
=\frac{ 1}{2^{N-\gamma} }\begin{pmatrix} N-\gamma \\ i- \gamma \end{pmatrix} \kappa^{(\Gamma)}(\ket{\phi_0}), \nonumber 
\end{equation}
where we have introduced the notation $\kappa^{(\Gamma)}(\ket{\phi_0}):=\sum_{a \neq b} r_a^2 r_b^2\prod_{i \in \Gamma}(1-\delta_{a_i b_i}) \prod_{i \notin \Gamma} \delta_{a_i b_i}$.
Thus $\Pi_{\infty}^{(\Gamma)}(\min)$ is given by
\begin{align}
\Pi_{\infty}^{(\Gamma)}(\min) &=\min_{i \in \{ \gamma, \cdots, N \}}
\frac{ 1}{2^{N-\gamma} }\begin{pmatrix} N-\gamma \\ i- \gamma \end{pmatrix} \kappa^{(\Gamma)}(\ket{\phi_0}) \nonumber  \\
&=\frac{1}{2^{N-\gamma} } \kappa^{(\Gamma)}(\ket{\phi_0}). \nonumber 
\end{align}

Let $\rho^{(\Gamma)}$ be the expression defined by Eq.~\eqref{InverseGap} for the Markov chain $\mathcal{\tilde{M}}_{\Gamma}$.
An upper bound of $\rho^{(\Gamma)}$ for $\mathcal{\tilde{M}}_{\Gamma}$ is then given by 
\begin{equation}
\rho^{(\Gamma)} \leq  \frac{\max_{e \in E}\sum_{\begin{subarray}{c} i,j \\ \mathcal{E}_{ij} \ni e \end{subarray}} \Pi^{(\Gamma)}_\infty(i) \Pi^{(\Gamma)}_\infty(j)}{\min_{e \in E}Q^{(\Gamma)}(e)}. \nonumber 
\end{equation}
Since the graph of the Markov chain $\tilde{\mathcal{M}_{\Gamma}}$ is linear, as shown in Fig.~\ref{newMarkov},
an upper bound on the maximum of $\sum_{\begin{subarray}{c} i,j \\ \mathcal{E}_{ij} \ni e \end{subarray}} \Pi^{(\Gamma)}_\infty(i) \Pi^{(\Gamma)}_\infty(j)$ is given by
\begin{align}
&\sum_{\begin{subarray}{c} i,j \\ \mathcal{E}_{ij} \ni e \end{subarray}} \Pi^{(\Gamma)}_\infty(i) \Pi^{(\Gamma)}_\infty(j)  \notag \\
&= \left(\frac{\kappa^{(\Gamma)} (\ket{\phi_0})}{2^{N-\gamma}} \right)^2 \max_{i \in \{1, \cdots N\}} \sum_{x=1}^{i} (\begin{subarray}{c} N-\gamma \\ x-\gamma \end{subarray}) \sum_{y=i+1}^{N} (\begin{subarray}{c} N-\gamma \\ y-\gamma \end{subarray}) \nonumber \\
&\leq \left(\frac{\kappa^{(\Gamma)} (\ket{\phi_0})}{2^{N-\gamma}} \right)^2 \max_{i \in \{1, \cdots N\}} \sum_{x=0}^{N-\gamma} (\begin{subarray}{c} N-\gamma \\ x \end{subarray}) \sum_{y=0}^{N-\gamma} (\begin{subarray}{c} N-\gamma \\ y \end{subarray} \nonumber )\\
&=(\kappa^{(\Gamma)}(\ket{\phi_0}))^2. \nonumber 
\end{align}

On the other hand, the $\min_{e \in E}Q^{(\Gamma)}(e)$ factor can be computed as follows.
In the Markov chain $\mathcal{\tilde{M}}_{\Gamma}$, edges are of the form $(i,i+1)$ or $(i,i-1)$.
By the symmetry of the linear graph, one sees that $Q^{(\Gamma)}(N+\gamma-i,N+\gamma-i-1)=Q^{(\Gamma)}(i,i+1)$, 
and it is sufficient to consider the minimum of $Q^{(\Gamma)}(i,i+1)$, which is given by
\begin{equation}
\min_{i} Q^{(\Gamma)}(i,i+1) = \frac{2\gamma(N-\gamma)}{2^{N-\gamma} N (N-1)} \kappa^{(\Gamma)}(\ket{\phi_0}). \nonumber 
\end{equation}
Thus $\rho^{(\Gamma)}$ is bounded from above as
\begin{equation}
\rho^{(\Gamma)} \leq \frac{2^{N-\gamma} N (N-1)}{2\gamma(N-\gamma)} \kappa^{(\Gamma)}(\ket{\phi_0}). \nonumber 
\end{equation}

In order to achieve
\begin{equation}
\forall T> T_{\mathrm{mix}}(\epsilon), \biggl| \expectation{{E}_L^{(A)}(\ket{\phi_T})} -  \langle E_L^{(A)} \rangle_{\Upsilon_{\rm phase}^{\rm comp}} \biggr| < \epsilon \nonumber 
\end{equation}
for all $A$, it is sufficient for each Markov chain $\mathcal{\tilde{M}}_{\Gamma}$ to converge with 
error $\epsilon' := \epsilon/2^N$ since the linear entropy is the sum of the stationary distributions in $\mathcal{\tilde{M}}_{\Gamma}$ as shown by Eq.~\eqref{Goal}. 
Therefore, from Theorem \ref{MixingTime} and Corollary \ref{AbsoluteGap}, we obtain an upper bound on $T_{\mathrm{mix}}(\epsilon)$ given by 
\begin{multline}
T_{\mathrm{mix}}(\epsilon)  \leq \max_{\Gamma} \biggl[ \frac{ N (N-1)}{2\gamma(N-\gamma)} 2^{N-\gamma}\kappa^{(\Gamma)}(\ket{\phi_0})  \biggr]^2 \nonumber  \\
\times \biggl[N- \gamma- \log \bigl(\frac{\epsilon}{2^N} \cdot \kappa^{(\Gamma)}(\ket{\phi_0}) \bigr) \biggr].
\end{multline}
This is dominated by the factor $[2^{N-\gamma}\kappa^{(\Gamma)}(\ket{\phi_0})]^2$.  Thus, $\max_{\Gamma} \kappa^{(\Gamma)}(\ket{\phi_0})=O(2^{-N})$ is sufficient for $T_{\mathrm{mix}}(\epsilon)$ to be a polynomial in $N$.

This concludes the proof.

\section{Summary}
We have defined phase-random states as an ensemble of states with fixed amplitudes and with uniformly distributed phases in a fixed basis.
We have discussed their use for the realization of canonical distributions in statistical mechanics. 
We then derived a general formula for the average amount of entanglement of phase-random states. 
Applying these results, we have argued for the simulatability of time evolving states by a Hamiltonian dynamics, and have shown the difficulty of their simulation for semi-classical Hamiltonian systems by MPSs. 
Finally, we have proven that an ensemble of states that provides the same average entanglement of phase-random states can be generated efficiently by a phase-random circuit composed of relatively simple gates. 

We acknowledge V. Vedral for useful comments.
This work is supported by Project for Developing Innovation Systems of MEXT, Japan and JSPS by KAKENHI  (Grant No. 222812, No. 23540463 and 23240001).

\appendix

\section{Concentration of measure} \label{App:Concent}

In this appendix, we show that for the ensemble of phase-random states with equal-amplitudes in a separable basis, $\Upsilon_{\rm phase}^{\rm eq,sep}$,
the amount of entanglement is highly concentrated around the average.  
Formally, by defining $\Delta  E_L^{(A)}(\ket{\phi^{\rm eq,sep}}) := |E_L^{(A)}(\ket{\phi^{\rm eq,sep}}) - \langle E_L^{(A)} \rangle_{\Upsilon_{\rm phase}^{\rm eq,sep}}|$ where $\ket{\phi^{\rm eq, sep}} = 2^{-N/2} \sum_{n=1}^{2^N} e^{i \varphi_n} \ket{u_n^{\rm sep}} $, we prove that $\mathrm{Prob}[\Delta E_L^{(A)}(\ket{\phi^{\rm eq,sep}}) > 2/2^{N}+\epsilon] \leq \exp[- c \epsilon^4 2^N] $ where $c=1/(2^{11} \pi^2)$.

First, for two states in the ensemble $\Upsilon_{\rm phase}^{\rm eq,sep}$ denoted by $\ket{\phi}=2^{-N/2} \sum_a e^{i \phi_a} \ket{u_a^{\mathrm{sep}}}$ and $\ket{\phi'}=2^{-N/2} \sum_a e^{i  \phi'_a} \ket{u_a^{\mathrm{sep}}}$, let us define the distance $d(\phi,\phi')$ between them in the parameter space 
$[0, 2\pi)^{2^N}$
by
\begin{equation}
d(\phi,\phi')=\frac{1}{2\pi}\sum_a \frac{|\phi_a - \phi'_a|}{2^N}. \nonumber 
\end{equation}
Then, using the theorems in Appendix C of Ref.~\cite{LPSW2009}, we obtain the upper bound of the concentration function $\alpha_d(r)$ by
\begin{equation}
\alpha_d(r) \leq \exp[-\frac{r^2}{8} 2^N], \label{Concentration}
\end{equation}
where the concentration function $\alpha_d(r)$ implies that, for any subset $A \in [0, 2\pi)^{2^N}$ with measure $1/2$, 
its $r$-neighborhood $A_r$ with respect to the metric $d$ has measure at least $1-\alpha_d(r)$ \cite{L2001}.

Now, we evaluate the amount of the change in the parameter space necessary to change $\Delta E_L^{(A)}(\ket{\phi^{\rm eq,sep}})$ more than $\epsilon$,
which is obtained from the following proposition;
\begin{Proposition} \label{PropDist}
For $\ket{\phi}$ and $\ket{\phi'} \in \Upsilon_{\rm phase}^{\rm eq,sep}$, 
\begin{equation}
|\Delta E_L^{(A)}(\ket{\phi}) - \Delta E_L^{(A)}(\ket{\phi'})| \leq 4 \sqrt{\pi}\sqrt{d(\phi,\phi')}.
\label{PropDistEq}
\end{equation}
\end{Proposition}

\begin{Proof}
Using the notation $\hat{\phi}_A = \tr_{\bar{A}} \ketbra{\phi}{\phi}$,
we calculate
\begin{align}
&|\Delta E_L^{(A)}(\ket{\phi}) - \Delta E_L^{(A)}(\ket{\phi'})| \notag \\
 &\leq  |E_L^{(A)}(\ket{\phi}) - E_L^{(A)}(\ket{\phi'})| \notag \\
&=|\tr \hat{\phi}_A^2 - \tr \hat{\phi'}_A^2| \notag \\
&=|\tr ( \hat{\phi}_A - \hat{\phi'}_A )( \hat{\phi}_A+ \hat{\phi'}_A)| \notag \\ 
&\leq |\!|\hat{\phi}_A - \hat{\phi'}_A|\!|_2  |\! |\hat{\phi}_A+ \hat{\phi'}_A|\! |_2  \label{b} \\
&\leq 2 D_{HS}( \hat{\phi}_A , \hat{\phi'}_A ) \notag \\
&\leq 2 D_{HS}( \ketbra{\phi}{\phi} , \ketbra{\phi'}{\phi'} )  \label{c} \\
&\leq 2 |\ket{\phi}-\ket{\phi'} |,  \notag
\end{align}
where $|\!|A|\!|_2:=\sqrt{\tr A A^{\dagger }}$ is the Hilbert-Schmidt norm and $D_{HS}(A,B)=|\!|A-B|\!|_2$.
Inequalities  \eqref{b} and \eqref{c} are obtained using Cauchy-Schwartz and Kadison's inequalities \cite{K1952}, respectively.
Since $|\ket{\phi}-\ket{\phi'} | \leq \sqrt{4 \pi d (\phi, \phi')}$ \cite{LPSW2009}, we obtain Eq.~(\ref{PropDistEq}).
\end{Proof}

Hence, in order to change $\Delta E_L^{(A)}(\ket{\phi^{\rm eq,sep}})$ more than $\epsilon$, $d(\phi,\phi')$ must be changed more than $\frac{\epsilon^2}{16 \pi}$.
Combining this with the concentration of measure given by \eqref{Concentration},
we obtain
\begin{multline} \label{Concentration2}
\mathrm{Prob}[\Delta E_L^{(A)}(\ket{\phi^{\rm eq,sep}})> \mu_M(\Delta E_L^{(A)}(\ket{\phi^{\rm eq,sep}}))+\epsilon] \\
 \leq \exp[- c \epsilon^4 2^N],
\end{multline}
where $\mu_M$ represents the median and $c=1/(2^{11} \pi^2)$.
By using Markov's inequality and the convexity of $\sqrt{x}$, the median is bounded from above such that
\begin{align}
\mu_M & (\Delta E_L^{(A)}(\ket{\phi^{\rm eq,sep}})) \leq 2 \langle \Delta E_L^{(A)} \rangle_{\Upsilon_{\rm phase}^{\rm eq,sep} } \nonumber  \\
&= 2 \biggl\langle \sqrt{(\Delta E_L^{(A)} )^2}  \biggr\rangle_{\Upsilon_{\rm phase}^{\rm eq,sep} } \nonumber  \\
& \leq 2 \sqrt{ \langle (\Delta E_L^{(A)} )^2}  \rangle_{\Upsilon_{\rm phase}^{\rm eq,sep} }  \nonumber  \\
&= 2 \sigma_{E_L^{(A)}}, \nonumber 
\end{align}
where $\sigma_{E_L^{(A)}}$ is the standard deviation of $E_L^{(A)}$.
Since the standard deviation $\sigma_{E_L^{(A)}}$ for ${\Upsilon_{\rm phase}^{\rm eq,sep} }$
can be directly calculated and is upper bounded by $2^{-N}$, we obtain
\begin{multline}
\mathrm{Prob}\biggl[\Delta E_L^{(A)}(\ket{\phi^{\rm eq,sep}})> \frac{2}{2^N}+\epsilon \biggr] \leq \exp[- c \epsilon^4 2^N]. \nonumber 
\end{multline}

\end{document}